\documentclass[usenatbib]{mnras}

\usepackage{amsmath}
\usepackage{siunitx}
\usepackage{xfrac}
\usepackage{natbib}
\usepackage{xcolor}
\usepackage{graphicx}
\usepackage{booktabs}
\usepackage{csquotes}
\usepackage{hyperref}            
            
\hypersetup{pdfpagemode = {UseNone},
            pdftitle = {Viscous-CBD-II},
            pdfcreator = {\LaTeX},
            pdfproducer = {pdfeTeX-0.\the\pdftexversion\pdftexrevision},
            pdfauthor = {Anna Penzlin, Wilhelm Kley},
            pdfsubject = {},
            pdfview = {FitH},
            pdfstartview = {FitH},
            colorlinks = {true},
            linkcolor = [rgb]{0,0.35,0.7},
            citecolor = [rgb]{0,0.35,0.7},
            filecolor = [rgb]{0.61,0,0},
            urlcolor = [rgb]{0.61,0,0},
           }
           
\usepackage{bm}
\renewcommand{\vec}[1]{\bm{\mathrm{#1}}}

\colorlet{darkgreen}{green!60!black}

\newcommand{\ab}{a_\mathrm{bin}}
\newcommand{\eb}{e_\mathrm{bin}}
\newcommand{\Tb}{T_\mathrm{bin}}

\newcommand{\beq}{\begin{equation}}
\newcommand{\eeq}{\end{equation}}

\title[II - Disc effects on the binary orbit]{ Viscous circumbinary protoplanetary discs - II.
Disc effects on the binary orbit}

\author[Penzlin et al.]{
Anna~B.T. Penzlin,$^{1}$\thanks{E-mail: a.penzlin@imperial.ac.uk}
Richard~A. Booth$^{2}$,
Richard~P. Nelson$^{3}$,		
Christoph~M.~Sch\"afer$^{4}$,
Wilhelm Kley\thanks{Passed away, but is included for his significant contribution.}
\\
$^{1}$Astrophysics Group, Department of Physics, Imperial College London, Prince Consort Rd, London, SW7 2AZ, UK \\
$^{2}$School of Physics and Astronomy, University of Leeds, Leeds LS2 9JT, UK\\
$^{3}$Astronomy Unit, Department of Physics \& Astronomy, Queen Mary University of London, London E1 4NS, UK\\
$^{4}$Institut f\"ur Astronomie und Astrophysik, Universität T\"ubingen,
Auf der Morgenstelle 10, 72076 T\"ubingen, Germany
}

\date{Accepted XXX. Received YYY; in original form ZZZ}

\pubyear{2024}

\begin{document}
\label{firstpage}
\pagerange{\pageref{firstpage}--\pageref{lastpage}}
\maketitle

\begin{abstract}
{
More than half of all stars are part of binaries, and many form in a common circumbinary disc. The interaction with the binary shapes the disc to feature a large eccentric inner cavity and spirals in the inner disc.
The shape of the cavities is linked to binary and disc properties like viscosity and scale height, and the disc and cavity shape influences the orbital evolution of the binary stars. 

This is the second part of the study in which we use 2D hydrodynamic long-term simulations for a range of viscous parameters relevant to protoplanetary discs to understand the interaction between young stars and the circumbinary disc.
The long-term simulations allow us to study how disc shape and exchange of mass, momentum and energy between binary and disc depend on the precession angle between disc and binary orbit on time scales of thousands of binary orbits.

We find a considerable, periodic interaction between the precession of the disc and the binary eccentricity that can significantly exceed the precession-averaged change in eccentricity.
We further confirm that thin discs ($H/R<0.05$) lead to shrinking binary orbits, also in the regime of low viscosity, $\alpha=10^{-3}$.
In general, the disc can excite eccentricity in binaries with initial eccentricities in the range of $\eb=0.05-0.4$. In most cases, the terms aiding shrinking or expansion and circularisation or excitation are nearly balanced, and the evolution of the binary semi-major axis and eccentricity will be sensitive to the ratio of mass accretion between the secondary and primary components. 
}\end{abstract}

\begin{keywords}
Hydrodynamics -- Binaries -- Accretion disc -- Protoplanetary discs
\end{keywords}

\section{Introduction}

In the investigation of star systems, the importance of binary stars is often underrated. Most stars above a solar mass do not form alone, and more than a quarter of all stars exist in pairs, even down to M-stars \citep{2022PPVII}. \cite{2020Whelan} found that binary stars in the APOGEE survey with orbits between 50-1000 days show a scatter in the binary eccentricity with peak values of $\eb=0.2-0.3$. As those stars likely formed inside a common circumbinary disc, this raises the question of what role the interaction of the disc with the binary stars played in excitation of the binary orbits.

Similarly, disc-binary interactions govern the accretion discs around supermassive binary black holes (SMBH) \citep{2023Lai}, where the interaction between disc and binary can significantly change the orbit of the binary, causing it to either shrink or expand. 
While the disc was traditionally thought to aid in binary merging, the studies of \cite{2017Miranda} and \cite{2019Munoz} showed that a thick, viscous disc can cause the binary to expand instead. Since then, numerous works have been dedicated to finding the limiting parameter values between expansion and shrinkage, given the binary eccentricity and mass ratios and disc viscosity and thickness.
\cite{2020Tiede}, \cite{2022Penzlin}, \cite{2022Franchini} and \cite{2022Dittmann} showed that discs with aspect ratios below 0.05 cause circular binary orbits to shrink. \cite{2023siwekI} found that the binary-disc interaction leads to eccentricity excitation of moderately eccentric binary orbits.

Recent SMBH studies \citep{2019Munoz, 2020Munoz, 2020Duffell, 2021Dorazio, 2021Tiede, 2022Dittmann, 2023siwekII, 2023siwekI} invoke viscosities equivalent to a viscous $\alpha\sim$0.3--0.01. This is orders of magnitude above the level expected for a protoplanetary disc(PPD) ($\alpha\sim 10^{-2}$--$10^{-4}$) \citep[e.g.][]{2018Dsharp6}. 
Thus, the regime of circumbinary PPDs has not yet been explored.

Decreasing the viscosity leads to long convergence times \citep{2018Thun,2020Hirsh} and makes the investigation more costly.  Also, the period of precession of the disc around the binaries acts on long time scales of thousands of orbits and, as we show in this work, the impact of the disc on the binary varies with the precession phase. Long-term simulations are thus needed to build an accurate picture of how the binary evolves.

In this study, we investigate the parameter space of viscous $\alpha$ and aspect ratio of protoplanetary discs to understand the dynamics of the disc, the interaction with the binary and the resulting binary orbital evolution. We use the 2D hydrodynamic simulation described in \cite{2024Penzlin}(PBN1), evolving each disc for $>30$k binary orbits ($\Tb$). Since the stars are not included in the domain, we use the angular momentum change, mass accretion and the energy change of the disc to investigate the change in the orbital parameters, as described in Sec.~\ref{sec:theory}. We will briefly reintroduce our simulation set in Sec.~\ref{sec:sims}. Sec.~\ref{sec:J_prec} shows the momentum contributions and different time scales relevant for understanding the overall orbit change on a standard example and Sec.~\ref{sec:e_bin_grav} highlights the role that the precession plays for the torques in the disc.
In Sec.~\ref{sec:range} the effect of the range of protoplanetary discs on the protostellar orbital evolution is shown and then discussed in Sec.~\ref{sec:discuss}.

\section{Theory of the orbital evolution}\label{sec:theory}
 
The binary's orbital evolution and disc dynamics are linked by the exchange of angular momentum and energy,
and changes in the semi-major axis and eccentricity of the binary result from momentum transfer, as described in \cite{2017Miranda}, for example.
	
Following previous descriptions we use the angular momentum change \citep[e.g.][]{2017Miranda, 2019Munoz, 2020Tiede, 2022Dittmann}, $\dot{J}_\mathrm{bin}$, and work done on the binary, $\dot{E}_\mathrm{bin}$, to predict the orbital elements of the binaries. 
Changes in gravitational energy and mass can lead to changes in the binary semi-major axis.
Using the ratio of $M_2/M_1=q$, and the mass accretion ratio, $\dot{M}_2/\dot{M}_1=\lambda$, the energy is given by
\begin{equation}
	E_\mathrm{bin} = -G \frac{M_\mathrm{bin}^2}{2 a_\mathrm{bin}} \frac{q}{(q+1)^2}
\end{equation}
Thereby the relative change in energy is given by:
\begin{align}
	\frac{\dot{E}_\mathrm{bin}}{E_\mathrm{bin}} & = 2\frac{\dot{M}_\mathrm{bin}}{M_\mathrm{bin}} - \frac{\dot{a}_\mathrm{bin}}{a_\mathrm{bin}} + \frac{\dot{q}}{q} \frac{1-q}{1+q} \nonumber \\
	& = \frac{\dot{M}_\mathrm{bin}}{M_\mathrm{bin}}\left( 2 + \frac{1-q}{1+\lambda} ( \frac{\lambda}{q} - 1)\right) - \frac{\dot{a}_\mathrm{bin}}{a_\mathrm{bin}} \label{eq:full_E},
\end{align}
and the semi-major axis evolution can be evaluated via:
\begin{align}
	\frac{\dot{a}_\mathrm{bin}}{a_\mathrm{bin}}= 
	 \frac{\dot{M}_\mathrm{bin}}{M_\mathrm{bin}}( 2 + \xi(q,\lambda)) -
	 \frac{\dot{E}_\mathrm{bin}}{E_\mathrm{bin}} \label{eq:full_a}
\end{align}
	
The possible values for $\xi(q,\lambda)$ are limited to be
$q-1<\xi<1/q-1$, for $\lambda=(0,\infty)$. This means the mass accretion term will always be positive, aiding expansion of the binary orbits. This means when the binary accretes material without changing the current velocity of the stars or the total energy of the system, the now more massive stars must be on a wider stable Keplerian orbit.

	\begin{figure}
		\centering
		\resizebox{\hsize}{!}{\includegraphics{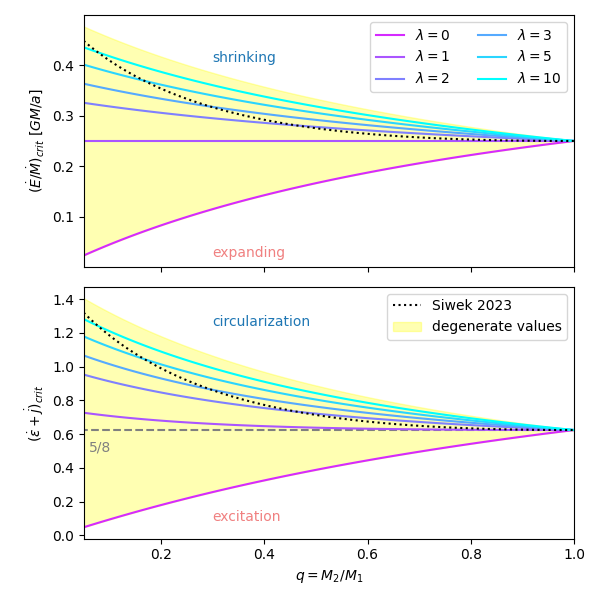}}
		\caption{Possible critical values of the orbit behaviour shown in Eq.~\ref{eq:crit_e} \& \ref{eq:crit_l} depending on mass and mass accretion ratio $q\&\lambda$. The top panel shows the critical value relevant for the semi-major axis evolution and the bottom for the eccentricity evolution. In the yellow region, both trends are possible depending on the $\lambda$ value.}
		\label{fig:xi}
	\end{figure}
Many previous studies have demonstrated that the secondary accretes more than the primary, such that the mass accretion ratio $\lambda>1$ \citep{2014Farris,2019Munoz,2021Dittmann,2023siwekI}.
In that case, the possible values are limited to $\xi>(1-q)^2/2q$, further increasing the expanding effect that mass accretion will have on the binary orbit.

Since we use $q=0.26$ for all models here, for the binary orbit to shrink the normalized power, $\dot{E}_\mathrm{bin}/E_\mathrm{bin}$, must be a factor $1.74 - 4.85$ larger than the normalized mass accretion rate, $\dot{M}_\mathrm{bin}/M_\mathrm{bin}$.
If we assume the relation between mass and mass accretion ratio in \cite{2023siwekI} ($\lambda\sim q^{-0.9}$), the factor is $>2$.
However, we note that there are studies that point to different values of $\lambda$, and even show accretion on to the more massive object is favoured object \citep{2014Gold}.

To find a critical value of the changes in mass and energy we can rearrange Eq.~\ref{eq:full_a} to the following
\begin{equation}
    \left(\frac{\dot{E}_\mathrm{bin}}{\dot{M}_\mathrm{bin}} \right)_\mathrm{crit} = - G \frac{M_\mathrm{bin}}{a_\mathrm{bin}}\cdot \frac{q}{2(q+1)^2} ( 2 + \xi(q,\lambda)). \label{eq:crit_e}
\end{equation}

Normalizing this in terms of the energy of a circular orbit this becomes

\begin{equation}
    \dot{\epsilon}_\mathrm{crit}= - \left(\frac{\dot{E}_\mathrm{bin}}{\dot{M}_\mathrm{bin}}\right)_\mathrm{crit} \frac{\ab}{G M_\mathrm{bin}} =  \frac{q}{(q+1)^2} ( 1 + \frac{1}{2}\xi(q,\lambda)).\label{eq:crit_e}
\end{equation}

The range of values for $q$ and $\lambda$ are shown in the first panel of Fig.~\ref{fig:xi}.
Using the $\lambda\sim q^{-0.9}$ in \cite{2023siwekI}, we find for our case of $q=0.26$ that the critical value of $\dot{\epsilon} \sim 0.33$.

Now, to understand the change in eccentricity of the binaries, we consider the change in angular momentum $\dot{J}_\mathrm{bin}$. Considering Eq.~\ref{eq:full_a}, the angular momentum and angular momentum changes of the binary star is given by:
	\begin{align}
	J_\mathrm{bin} &=\frac{q}{(1+q)^2} \sqrt{G M_\mathrm{bin}^3 a_\mathrm{bin}(1-e_\mathrm{bin}^2)} \\ 
	\frac{\dot{J}_\mathrm{bin}}{J_\mathrm{bin}} & = \frac{1}{2}\frac{\dot{a}_\mathrm{bin}}{a_\mathrm{bin}}+ \frac{3}{2}\frac{\dot{M}_\mathrm{bin}}{M_\mathrm{bin}}- \frac{\dot{e}_\mathrm{bin}}{e_\mathrm{bin}} \frac{e_\mathrm{bin}^2}{1-e_\mathrm{bin}^2}+\frac{\dot{q}}{q} \frac{1-q}{1+q} \nonumber \\ \nonumber
	& = \frac{1}{2}\frac{\dot{M}_\mathrm{bin}}{M_\mathrm{bin}} \left[ 5 + 3 \frac{1-q}{1+\lambda} ( \frac{\lambda}{q} - 1) \right]\\
	&\quad - \frac{1}{2}\frac{\dot{E}_\mathrm{bin}}{E_\mathrm{bin}}
	- \frac{\dot{e}_\mathrm{bin}}{e_\mathrm{bin}} \frac{e_\mathrm{bin}^2}{1-e_\mathrm{bin}^2} \label{eq:full_J}
	\end{align}
Thereby, the change of eccentricity is given by
	\begin{equation}
	\frac{\dot{e}_\mathrm{bin}}{e_\mathrm{bin}}	= \left( \frac{\dot{M}_\mathrm{bin}}{M_\mathrm{bin}} \left[ \frac{5}{2} + \frac{3}{2} \xi \right] - \left[\frac{1}{2}\frac{\dot{E}_\mathrm{bin}}{E_\mathrm{bin}} + 	\frac{\dot{J}_\mathrm{bin}}{J_\mathrm{bin}}\right]\right)\frac{1-\eb^2}{\eb^2}. \label{eq:full_e}
	\end{equation}
	
The mass accretion term will again always be positive, leading to eccentricity excitation. Considering the $\xi$ values above, the binary orbit will circularize when the normalized combined power and momentum term is a factor $2.11 - 6.78$ larger than the normalized mass accretion term, $\dot{M}_\mathrm{bin}/M_\mathrm{bin}$. If the mass accretion is in favor of the secondary ($\lambda\geq1$), the factor will exceed $4$.

Similarly to the specific energy term, $\dot{\epsilon}$, we can define a angular momentum term $\dot{j}$ normalized to the mass accretion rate,

\begin{equation}
    \dot{j} = \frac{\dot{J}_\mathrm{bin}}{\dot{M}_\mathrm{bin}}(G M_\mathrm{bin} a_\mathrm{bin}(1-e_\mathrm{bin}^2))^{-1/2}
\end{equation}

We can find the critical condition between circularization and excitation by multiplying with the specific mass accretion ratio to separate all terms that purely depend on $q$ and $\lambda$,
\begin{equation}
(\dot{\epsilon}+\dot{j})_\mathrm{crit} = \left[ \frac{5}{2} + \frac{3}{2} \xi \right] \frac{q}{(1+q)^2}\  \label{eq:crit_l}.
\end{equation}

The lower panel of Fig.~\ref{fig:xi} plots the total modulation due to prefactor of $\dot{M}/M$ and the mass ratio dependence $q/(q+1)^2$ factored in the normalized energy and angular momentum change $(\dot{\epsilon}+\dot{j})_\mathrm{crit}$. 
This figure includes all possible limiting values between the full accretion only on the primary or secondary.
If the derivatives exceed these values growth and excitation of orbits are possible.
The value relevant for $q=0.26$ is $(\dot{\epsilon}+\dot{j})_\mathrm{crit} \sim 0.9$ using the estimate in \cite{2023siwekI}, but can be between 0.2 to 1.2 depending on $\lambda$.
 
To directly understand the amount of energy and angular momentum exchanged between disc and binary, we look at two mechanisms of exchange, gravitational interaction and advected contributions.
The gravitational torques $\tau$ and the energy change $\dot{E}_\mathrm{grav}$ the binary experiences due to the disc are calculated using the potential between cell k and either binary star i, $\Phi_i$.
	\begin{align}
	\tau(t) &=  \int_r^{r_{out}}\frac{\mathrm{d} \tau}{\mathrm{d} r}(t)\mathrm{d} r  \label{eq:torq}\\
	\frac{\mathrm{d} \tau}{\mathrm{d}r}(t) &=  \oint_\phi \sum_i \Sigma(t) \frac{\partial\Phi_i}{\partial\phi}(t) r_i \mathrm{d}\phi \\
	\dot{E}_\mathrm{grav}(t) &= - \int_r^{r_{out}}\oint_\phi \sum_i \Sigma(t) \nabla \Phi_i (t) \cdot \vec{v}_i(t) \mathrm{d}\phi \mathrm{d}r \label{eq:power}\\
	\mathrm{ with} \quad &\Phi_{i=1,2}(t) = - \frac{Gm_i}{(r_i^2 + r_k^2 - r_i r_k \cos(\phi_k-\phi_i))^{1/2}}
	\end{align}
	
To evaluate the accretion behaviour we look at the mass flow $ \dot{M}_\mathrm{acc}$ and advective angular momentum flow $\dot{J}_\mathrm{acc}$ through the inner boundary $r_\text{ev} = 1 \,\mathrm{a_{bin}}$. With this the accretion can be calculated by the following:
	\begin{align}
	\dot{M}_\mathrm{acc}(t) & = - \oint_\phi \Sigma(t) r v_{r}(t) d\phi \bigg|_{r_\text{ev}} \\
	\dot{J}_\mathrm{acc}(t) & = - \oint_\phi \Sigma(t) r^3 v_{r}(t) \Omega(t) d\phi \bigg|_{r_\text{ev}}
	\end{align}
	
Thereby, the change in binary angular momentum is given by $\dot{J}_\mathrm{bin} = (\dot{J}_\mathrm{acc} + \tau)$.

The binary and disc exchange mass, angular momentum, and energy. By measuring the changes in the disc's mass, momentum, and energy, we can compute the binary's orbital evolution due to these contributions. These measurements are possible even if the binary is not in the domain and reacting to the disc, as the binary's orbital evolution happens on long timescales.

The viscous torque needs to be taken into account as well to complete the analysis of the angular momentum exchange. However, it is more than four orders of magnitude smaller than all other effects at the inner boundary, so we neglect it.

\section{Simulation setup}\label{sec:sims}
	
    \begin{figure}
		\centering
		\resizebox{\hsize}{!}{\includegraphics{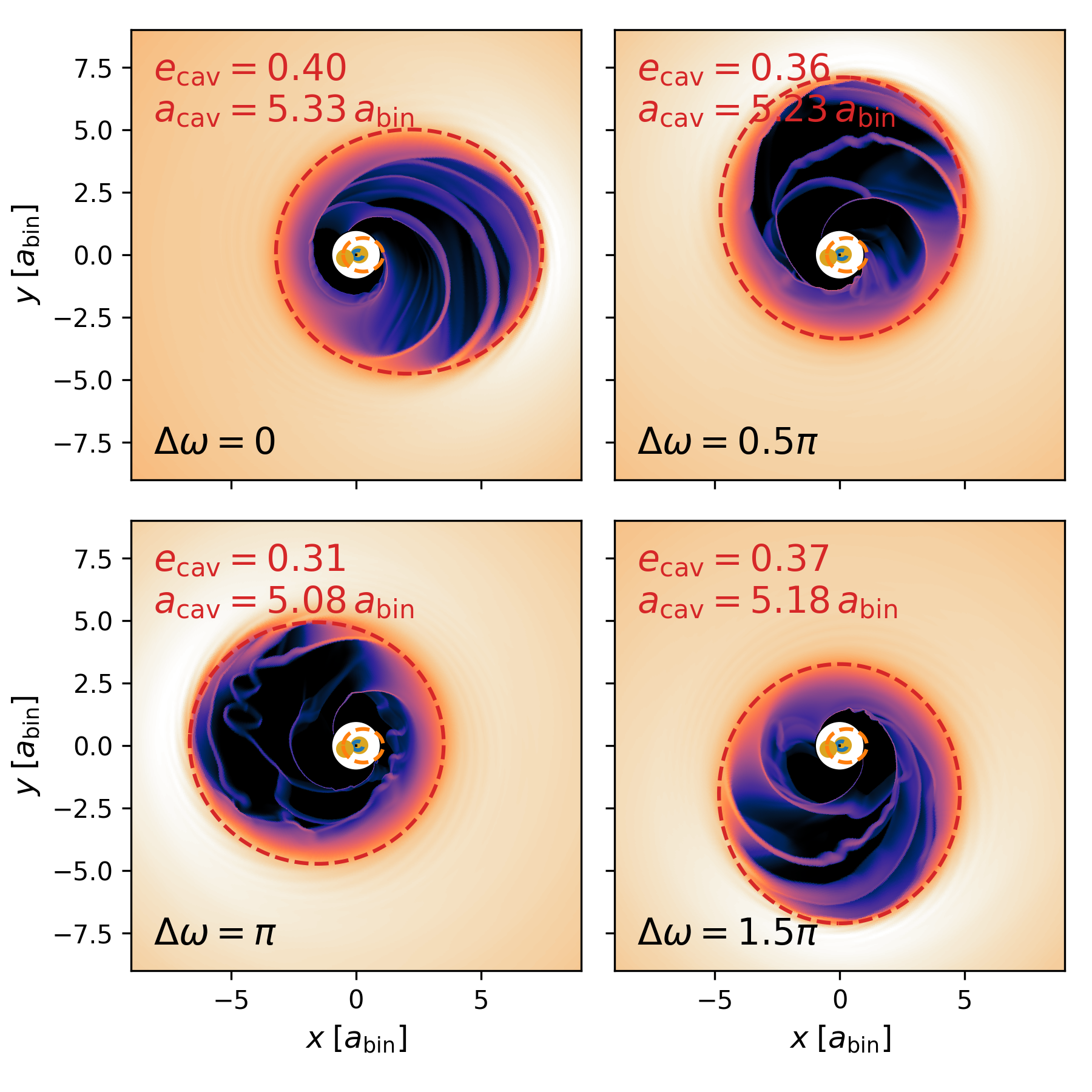}}
		\caption{The logarithmic surface density structure of the circumbinary disc at different precession angles. The binary eccentricity is 0.4 in this setup and scales are in binary semi-major axis $\ab$.
        The orbits of the binary stars (orange and blue dashed line) are fixed the cavity ellipse is traced by the outer red dashed line with orbit parameters listed in the top left corner and the angle between binary and cavity pericentre $\Delta\omega$ is noted in the bottom left corner.}
		\label{fig:2d-ex}
	\end{figure}

The simulation setup used here has been described in detail in the first paper, PBN1. We use the same locally isothermal 2D hydrodynamic PLUTO simulations, restarting the simulations from a converged state of the disc at $\geq 3\times 10^4~\Tb$ and continuing them while running the analysis of all momentum and energy contributions for another $\geq 10^4~\Tb$.
In this time, the most viscous disc will reach the viscous time scale at $4~\ab$, while the disc with the lowest viscosity runs to a only few percent of the viscous time at this radius. PBN1 discuss the fact that the disc evolution caused by the binary does not act on viscous time scales but leads to eccentricity excitation throughout the disc on a much faster timescale of a few thousand orbits (see also the Appendix in PBN1).

As in the previous study, we use the mass ratio $q=M_2/M_1=0.26$. For a fiducial disc case, the binary eccentricity is $\eb=0.2$, the aspect ratio $h=0.05$, and the viscous $\alpha=10^{-3}$.
In addition, we investigate a range of aspect ratio and viscous $\alpha$ for binary eccentricities, $\eb= [0, 0.4]$. The aspect ratio takes values from 0.03 to 0.1 in steps of 0.01 and the $\alpha$-values include [$10^{-4}, 2.5\times 10^{-4}, 10^{-3}, 2.5\times 10^{-3}, 10^{-2}$]. The domain runs from $1~\ab$ to $40~\ab$ with $N_r\times N_\phi = 684\times584$ cells. 
This large computation domain allows the disc to develop a globally eccentric structure that is not impacted by the outer boundary condition.
 The discs have a constant aspect ratio $H/R=h=$const. and receive a continuous mass according to the initial profile in the outer damping zone that covers the final 10\% of the domain. The initial surface density in PBN1 scaled with $\Sigma_\mathrm{ini} \propto R^{-1.5}$, which is not consistent with the profile of a steady-state viscous accretion disc. However, \cite{2020Munoz} showed that the shape of the inner disc and the mass normalized terms do not depend on the disc profile. We verified this by comparing the simulations presented in the main text with simulations using surface density profiles close to those expected from the viscous steady-state in Appendix~\ref{sec:app}.
The inner boundary adopts a diode-like outflow condition for the radial velocity. The gradient of azimuthal angular velocity and surface density at the inner boundary are set to zero.

An example of the converged structure of the circumbinary disc is given in Fig. \ref{fig:2d-ex}. The motion of the binary clears a wide, eccentric inner cavity over $\sim 10^4~\Tb$ with little dependency on the viscosity, as we have shown in PBN1. We measure the cavity as a 90\% radial drop of the surface density and find typical cavity sizes between $3.5 - 6~\ab$, with eccentricities between $e_\mathrm{cav}=0.1-0.45$. Gas in the inner disc moves on eccentric Keplerian orbits, which leads to a pressure maximum near the slow-moving apocenter location. 
The disc adopts a smoothly varying globally eccentric structure in time and space, with the region of highest density being confined to the apocenter. Further discussion of the features and structures of the disc can be found in PBN1.

The inner disc starts to precess on time scales of thousands of orbits due to the continued interaction with the binary. The precession timescale can be approximated by the massless three-body solution \citep{2004Moriwaki} at the eccentric orbit of the density maximum. However, the precise precession rate varies because hydrodynamic effects affect the eccentricity propagation into the disc. In particular, the pressure scale height of the disc can alter the region that rigidly precesses around the binary and increase the precession period up to 1.8 times the value of the theoretical solution for the period of the peak position(see PBN1). The cavity size and eccentricity oscillate with the disc precession period, affecting the interaction between binary and disc.

\section{Orbit and precession effects} \label{sec:J_prec}
	
	\begin{figure}
		\centering
		\resizebox{\hsize}{!}{\includegraphics{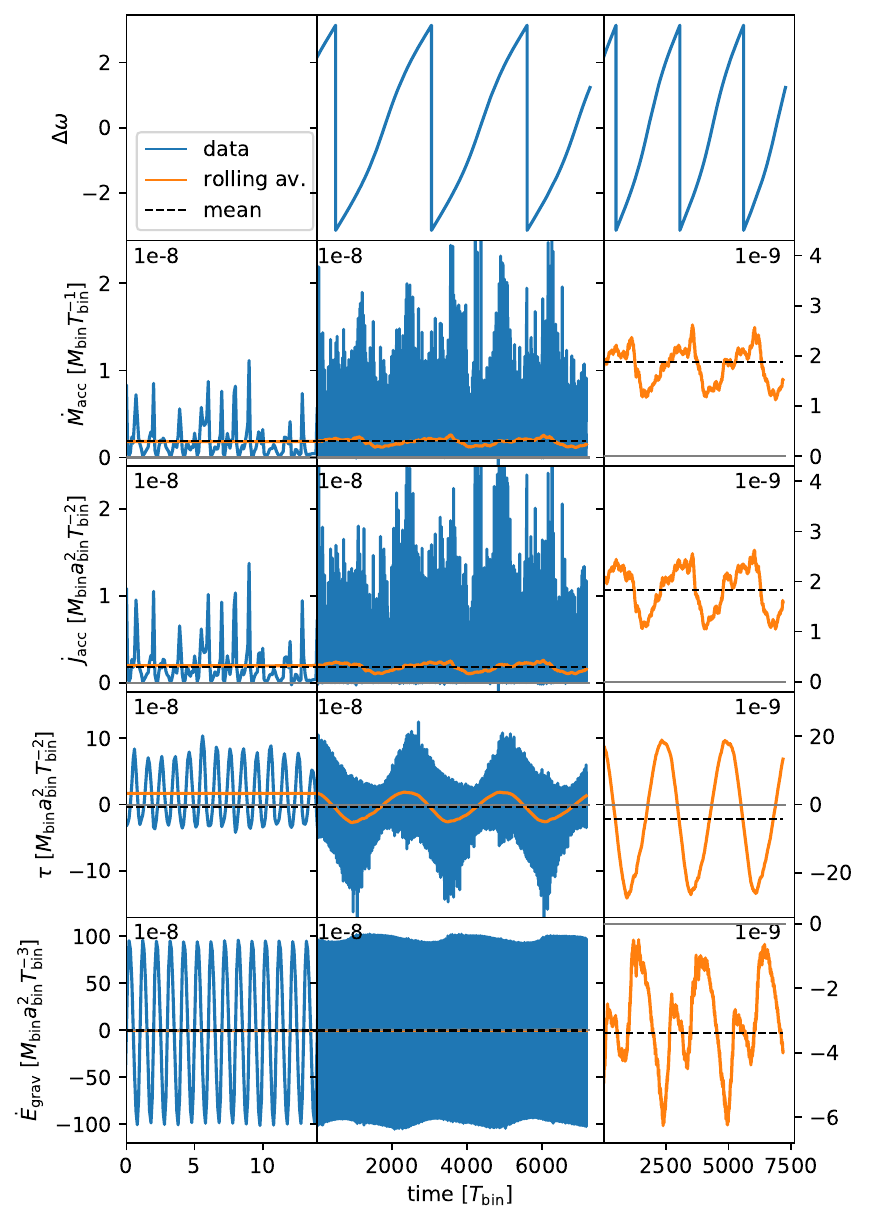}}
		\caption{Pericenter of the disc eccentricity vector (top), massflow(2nd) and advective angular momentum flow (3rd) through the inner boundary accreted by the binary, gravitational torque (4th) and energy change caused by the disc on the binary are plotted over $15 ~\Tb$ (left) and $7000~\Tb$. The orange lines mark a rolling average over $100~\Tb$ and the black lines mark the absolute averages.}
		\label{fig:evo-time}
	\end{figure}
	
All of the contributions leading to the change in the binary orbit described in Sec.~\ref{sec:theory} can be analysed by considering the motion and distribution of the disc gas in the models.
For the fiducial model setup ($\eb=0.2; h=0.05; \alpha=10^{-3}$), Figure~\ref{fig:evo-time} shows the evolution of the angular momentum transport and the mass accretion and power on orbital and precession time scales through the inner boundary. 
	
All terms vary on orbital time scales, as shown in the left panels. The torque, $\tau$, and gravitational power, $\dot{E}$, strongly oscillate every orbit, while advected mass $\dot{M}$ and angular momentum $\dot{J}_\mathrm{acc}$ peak once per orbit with varying strength. 
Comparing the instantaneous contributions to their averages over $100~\Tb$ windows (middle panel), we see that the orbital variations in the gravitational terms and the peaks from the advection terms are an order of magnitude larger than the averaged values. 
The orbit-averaged terms show a further oscillatory behaviour on the precession timescale of the disc. 
The normalized mass and momentum accretion, $\dot{M}$ \& $\dot{J}_\mathrm{acc}$, are nearly identical in shape and their mean values are nearly identical. 
Thus, during the accretion, the orbital velocity of the accreted material depends little on the alignment of the disc with respect to the binary.
For the gravitational terms, the structure of the oscillation is visible before orbit averaging, with the biggest effect on the gravitational torque, $\tau$. Here, even the orbit-averaged terms change sign over the precession period, oscillating with an amplitude that is an order of magnitude larger than the mean value obtained by averaging the torque over a full precession period. This substantial variation in the angular momentum change will impact the evolution of the eccentricity (see Eq.~\ref{eq:full_e}) and make the short time-scale averages of the torque unreliable.

To illustrate the significance of the precession phase, we ran 14 simulations covering 50 orbits at different alignments during one precession period. Each run stored 20 steps per orbit, ensuring that the orbital periodicity is well sampled and does bias to the averaged data.
	
\begin{figure}
		\centering
		\resizebox{\hsize}{!}{\includegraphics{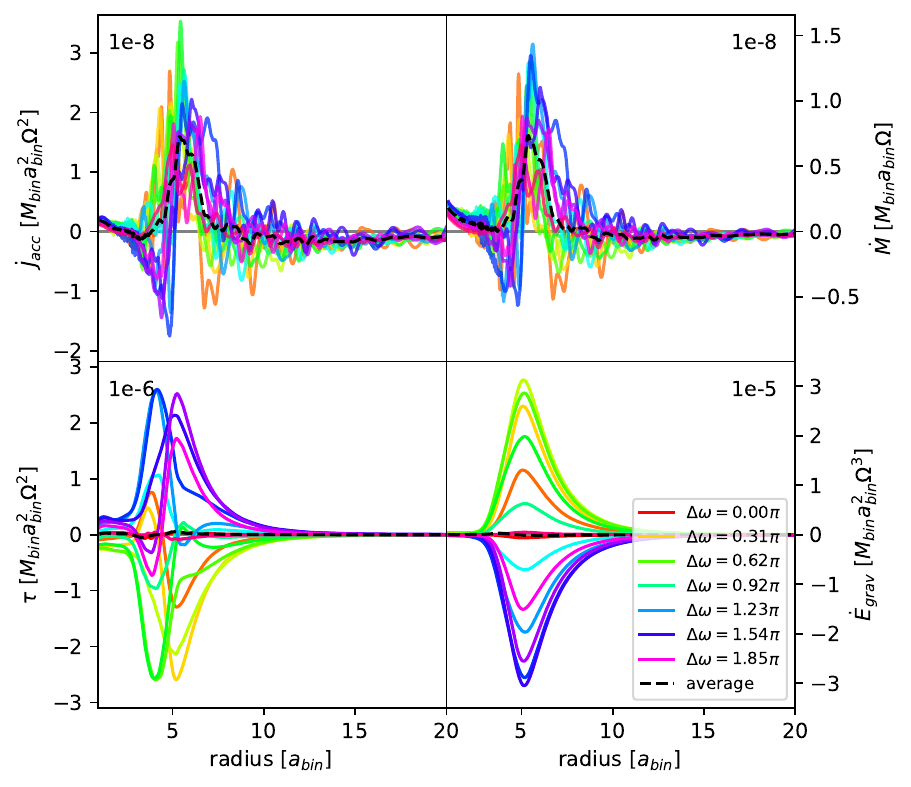}}
		\caption{Advective angular momentum flow(top left), massflow(top right), torque caused by the disc on the binary(bottom left) and energy change on the binary(bottom right) are plotted radially. The relative argument of periastron between binary orbit and disc $\Delta \omega$ of every second simulation is given in the legend.}
		\label{fig:precession}
	\end{figure}

Fig. \ref{fig:precession} shows how the radial profiles change over one precession period.
All contributions show how the maximum density ($r_\mathrm{peak}\sim6$) in the inner disc results in the largest effect, and the contributions reduce in the cavity ($a_\mathrm{cav}\sim4$).
While the general structure of advection through the disc remains the same over time, as seen in mass and angular momentum flux, $\dot{M}$ and $\dot{J}_{adv}$, the gravitational torque $\tau$ and power $\dot{E}_\mathrm{grav}$ depend on the precession phase. 
The mass and angular momentum flux also show some comparable trends, especially at the rim of the disc and the cavity edge between 3 and $6\,a_\mathrm{bin}$. As the surface density profile is steeper than a steady-state, gas near the inner cavity spreads ring-like towards both directions like in \cite{2020Munoz}(see also the Appendix \ref{sec:app}). However, the accretion is subject to unpredictable short-term variations depending on the local structure of spiral features and wave propagation and is, thereby, inherently noisy.
	
\cite{2020Tiede} studied the local origin of the
torque acting on the binary and examined how this varies with the orientation of the eccentric cavity wall.
They only considered equal mass binaries on fixed circular orbits, introducing a symmetry into the problem. In this work, we generalise the problem to consider non-equal mass binaries on eccentric orbits, showing how the changing orientation between the eccentric disc and the eccentric binary modulates the interaction. A key result of both studies is that the circumbinary disc plays an essential role in driving the orbital evolution of the binary.

	\begin{figure}
		\centering
		\resizebox{\hsize}{!}{\includegraphics{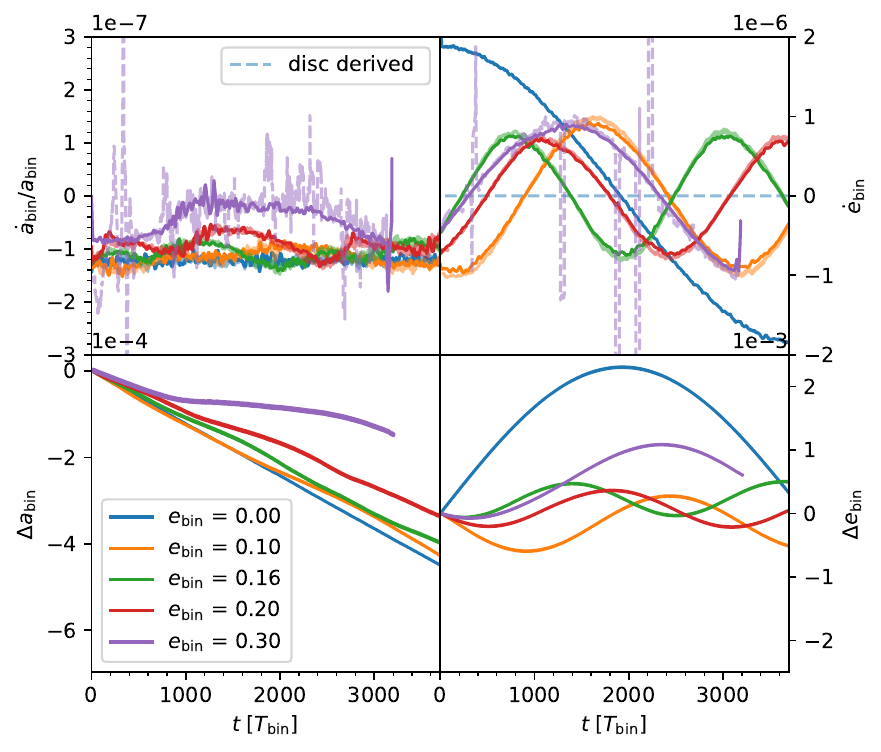}}
		\caption{The change of the semimajor axis and the eccentricity (darker lines) in a non-accreting scenerio and the calculation of the semimajor axis and eccentricity from the gravitational power and torque (dashed). The viscous $\alpha = 10^{-3}$ and the disc scale height is $h=0.05$}
		\label{fig:orbit-change}
	\end{figure}
	
 \section{Variations of the gravitational contributions} \label{sec:e_bin_grav}
	To study the influence of the gravitational terms on the binary orbit, 
	the fiducial simulation with 5 different initial binary eccentricities has been continued after $80\,000\, T_\mathrm{bin}$ with the gravitational disc feedback onto the binary stars included to simulate the change in orbit parameters.
	As the inner domain starts at $r_\mathrm{in}=1~\ab$, the binary accretion is not modelled and the simulation gives no insight how the mass accretion is distributed between the stars.
	Therefore, the $\dot{M}$ and $\dot{J}_{\rm adv}$ are neglected, but the gravitational interactions are captured by the disc feedback on the binary in the n-body interaction \citep[see also ][]{2018Thun}. A simple, accretion-less form of Eqs.~\ref{eq:full_E} and \ref{eq:full_J} can be compared to the binary elements. Under these assumptions, the binary evolves according to:
	\begin{align}
	\frac{\dot{E}_\mathrm{grav}}{E_\mathrm{grav}} &= -\frac{\dot{a}_\mathrm{bin}}{a_\mathrm{bin}}  \label{eq:simple_E_dot/E} \\
	\frac{\dot{e}_\mathrm{bin}}{\eb} &= -\left[\frac{\dot{J}_\mathrm{bin}}{J_\mathrm{bin}} + \frac{1}{2}\frac{\dot{E}}{E} \right] \frac{1-e_\mathrm{bin}^2}{2e_\mathrm{bin}^2}
	\label{eq:simple_J_dot/J}
	\end{align}
	
	Fig.~\ref{fig:orbit-change} compares the evolution of the binary orbit due to disc gravity to the gravitational change in orbit parameters calculated solely from the disc properties.
	However, it is important to note that energy conservation will also depend on the mass accretion and the accretion ratio and for more realistic models that include accretion using the full equations, Eqs.~\ref{eq:full_a} \& \ref{eq:full_e}.
	 
	The binary orbital evolution follows the parameter changes derived from the gravitational disc power and torque.
	The gravitational power leads to the binary orbit shrinking in these simulations. The change in semimajor axis, $\dot{a}_\mathrm{bin}$, varies periodically on the timescale of disc precession, and the torque-driven eccentricity change $\dot{e}_\mathrm{bin}$ is even more dramatic.
	The simulated binary eccentricity experiences a sine-like oscillation corresponding to the disc precession. This should be generally expected as the gravitational torque variations are an order of magnitude higher than other angular momentum transfer contributions shown in Fig.~\ref{fig:evo-time}. 
 For $\eb=0.2$, the gravitational torque already varies by more than the magnitude of the mean value and, therefore, changes signs, while for $\eb=0.4$, the variations reach an order of magnitude higher values than the mean torque. This is more severe for lower viscosities, where the mean torques are lower but the variations remain of the same strength.
 This long-term periodic behaviour poses a challenge for the study of the trends of binary orbit behaviour.
The eccentricity variability caused by the precession reaches values of $\eb\sim\pm10^{-3}$. Thereby, a model run for only $1000~\Tb$ can vastly misrepresent the trend of eccentricity change.
The contributions to the angular momentum change need to be considered on a time scale of multiple precession periods and the slow oscillatory behaviour of the torque must be taken into account when evaluating average values for the contributions to understand the general trends.
Long-term, temporally highly resolved simulations are needed to investigate the orbit dynamics directly. 

\begin{figure*}
	\centering
	\resizebox{\hsize}{!}{\includegraphics{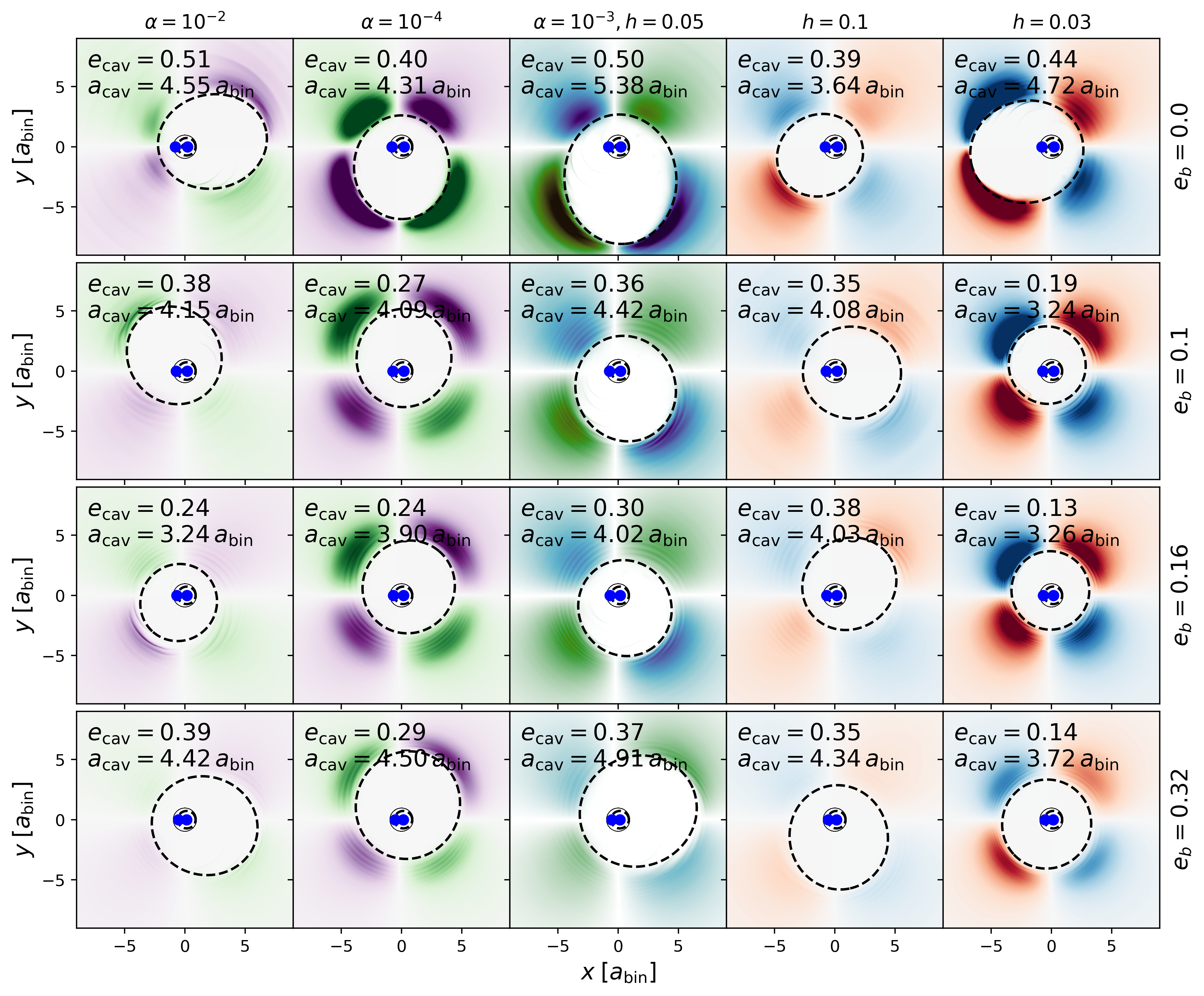}}
	\caption{The gravitational local torques of the simulations in PBN1. The torque is normalized to the total disc mass. Left two columns are the minimum and maximum values of $\alpha$ and the right two columns the same for the aspect ratio $h$. The mid column is are the fiducial parameters. Rows represent different $\eb=[0,0.1,0.16,0.32]$. The black dashed line marks the edge of the disc. All colour maps use the same linear scales centred on zero torque in white. The colour in the top left of each panel represents negative and the top right colour positive torque contributions.}
	\label{fig:maps_torq}
\end{figure*}
 
	\section{Effects of viscosity and aspect ratio}\label{sec:range}

Figure~\ref{fig:maps_torq} shows the gravitational torque between the local disc gas and the binary stars as derived in Eq. \ref{eq:torq} for the maximum and minimum values of $\alpha$ and $h$ and the fiducial disc parameters. As in Fig.~\ref{fig:precession}, this already shows that the density in the cavity is so low that it does not contribute a significant torque independent of the proximity throughout the whole investigated parameter space. 
An obvious feature of these maps is that the low viscosity case with $\alpha=10^{-4}$ and the h=0.03 both show large amplitudes in the magnitudes of the torques (see the second and fifth columns). This is in agreement with the results of \cite{2020Tiede}. However, it is important to noted that the net torque experienced by the binary is obtained by integrating around the orbit. And so the effect of the disc on the binary can not be be understood by simply inspecting Fig.~\ref{fig:maps_torq} due to cancellation effects.
As the discs beyond the cavity are eccentric, the total integrated torque will be strongly affected by the binary-to-disc alignment, as shown in Fig.~\ref{fig:precession}. 
Hence, to understand the averaged effect the torque analysis needs to be well resolved in time.
 
As the initial simulations in PBN1 did not include a runtime analysis of the angular momentum and energy exchange, the simulations have been continued from the evolved disc with analysis once every $0.1~\Tb$. An example of the final data is Fig.~\ref{fig:evo-time}.

To achieve the best possible averaged momentum and energy values for the terms depending on the gravitational potential, we averaged the data over multiple complete precession periods in the simulations. The statistical uncertainty of these contributions to the orbital evolution were estimated by considering how the precession-period averaged values vary from period to period. This was achieved by moving averages with a duration of one precession period and start points that are continuously shifted across a half period. The standard deviation of these shifted means is used as the noise in the signal-to-noise calculation.
For higher $\eb$, when the secondary comes close to the domain or even enters it, we exclude the inner most regions of the domain from the torque and energy calculation. We have used the apocentre of the secondary with an added margin of $0.3~\ab$ as the boundary. We checked that this choice does not affect the result, by comparing the results to those that exclude the region inside $1.5~\ab$ and found similar values for both.

With this averaged data, we can understand the direction of binary orbital evolution for the different viscous $\alpha$ (Fig.~\ref{fig:sur_a}) and aspect ratios (Fig.~\ref{fig:sur_h}).
To create a estimation of the continuous parameter space we use a ``linear" interpolation in between the simulation averages which are placed as dots in the maps. Linear creates a continuous linear transition between the nearest data points.

\subsection{Scale height}

\begin{figure}
	\centering
	\resizebox{\hsize}{!}{\includegraphics{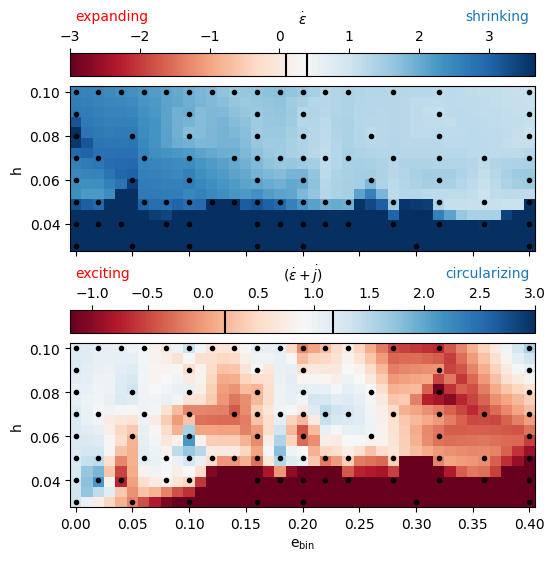}}
	\caption{
 Energy change rate normalized by the mass accretion rate(top) and normalized energy and angular momentum change (bottom) for different scale heights $h$. Black dots mark different simulations. Red values lead to expansion or excitation, blue values to shrinkage or circularization for $q=0.26$ with the $\lambda$-dependent range of critical values indicated with black lines (see Eq.~\ref{eq:crit_e} \& \ref{eq:crit_l}).}
	\label{fig:sur_h}
\end{figure}

Using the critical limits presented in Fig.~\ref{fig:xi}, we show the direction of orbital evolution for different aspect ratios and binary eccentricities in Fig.~\ref{fig:sur_h}.

The change in energy of the binary, $\dot{E}$, is generally negative, leading to shrinking orbits. As $\tau$ and $\dot{E}$ are both based on the gravitational potential their general trends are similar. The absolute value of $\dot{E}$ decreases beyond $\eb>0.2$, with the binaries entering the region of parameter space where the orbit may expand or shrink, depending on the mass ratio and mass accretion ratio. The white colour in Fig.~\ref{fig:sur_h} is centered on 0.33. This is equivalent to the limits between expansion and shrinkage for $q=0.26$ using \cite{2023siwekI}.
For thin discs, $\dot{E}$ significantly increases. This means binary orbits for thin discs with $h<0.05$ can only shrink.

To charaterise the eccentricity evolution, we can use the normalized energy-momentum term $(\dot{\epsilon}+\dot{j})$, as given in Eq.~\ref{eq:crit_l}. The bottom panel of Fig.~\ref{fig:sur_h} shows the magnitude of these terms centred on 0.9 which is the critical value according to \cite{2023siwekI}.
One clear trend for thin discs can be found. For $h<0.05$ the orbits of eccentric binaries are excited.
For discs thicker than $h=0.05$, the resulting 
level of angular momentum and energy exchange is close to the critical value at which the orbital evolution switches between gaining and losing eccentricity and the actual evolution depends on the mass accretion ratio.

\begin{figure}
	\centering
	\resizebox{\hsize}{!}{\includegraphics{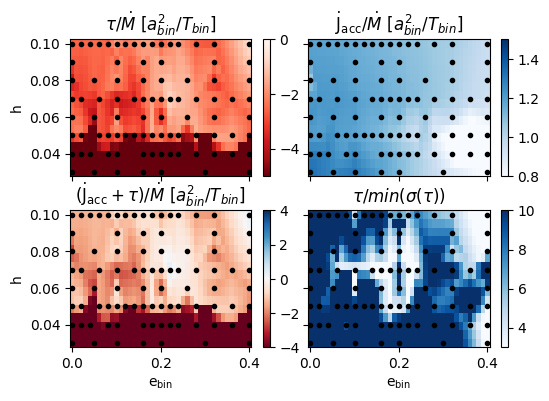}}
	\caption{The torque (top left) and the angular momentum change (bottom left) caused by the disc on the binary. The angular momentum flow normalized with the mass flow (top right) and torque signal relative to the deviation from the mean (bottom right) for different aspect ratios, $h$. Black dots mark different simulations.}
	\label{fig:sur_h_j}
\end{figure}

The contributions of the individual terms to the orbital evolution are shown in Fig~\ref{fig:sur_h_j}.
The two accretion-dependent terms, $\dot{M}$ and $\dot{J}_\mathrm{acc}$, can only add mass and momentum to the binary and must thereby be positive. As all disc dynamics is linked to the accretion processes, we use $\dot{M}$ to normalize the other contributions in Fig.~\ref{fig:sur_h_j}. Normalizing the other terms by $\dot{M}$ yields results near 1, indicating that all processes are comparable to the magnitude of the general viscous accretion in an accretion disc.
The two angular momentum changes $\tau$ and $\dot{J}_\mathrm{acc}$ are counteracting each other. The advected angular momentum increases the total angular momentum, while the gravitational torque $\tau$ mostly reduces angular momentum. 
$\dot{J}_\mathrm{acc}$ is closely linked to $\dot{M}$ such that $\dot{J}_\mathrm{acc}/\dot{M}$ only varies between [0.8, 1.3] throughout the whole parameter space. 
The $\dot{J}_\mathrm{acc}$ contribution decreases with increasing binary eccentricity, and this trend becomes a slightly more significant for lower aspect ratios.
The torque varies over a broader range than $\dot{J}_\mathrm{acc}$ and thereby dominates the overall angular momentum change, resulting in a near-universally negative total torque that contributes to the binary's eccentricity excitation.

\subsection{Viscous alpha}

\begin{figure}
		\centering
		\resizebox{\hsize}{!}{\includegraphics{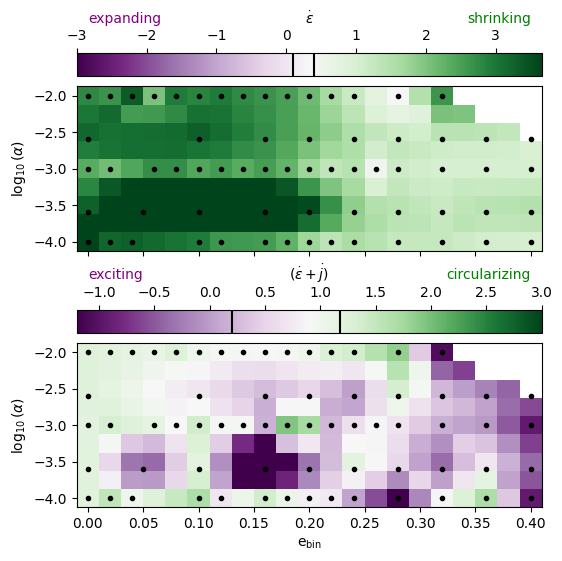}}
		\caption{Accretion weight energy change rate(top) and energy and angular momentum change rate (bottom) for different $\alpha$-values. 
    Black dots mark different simulations. Purple values lead to expansion or excitation, green values to shrinkage or circularization for $q=0.26$ with the $\lambda$-dependent range of critical values indicated with black lines (see Eq.~\ref{eq:crit_e} \& \ref{eq:crit_l}).}
		\label{fig:sur_a}
	\end{figure}

\begin{figure}
		\centering
		\resizebox{\hsize}{!}{\includegraphics{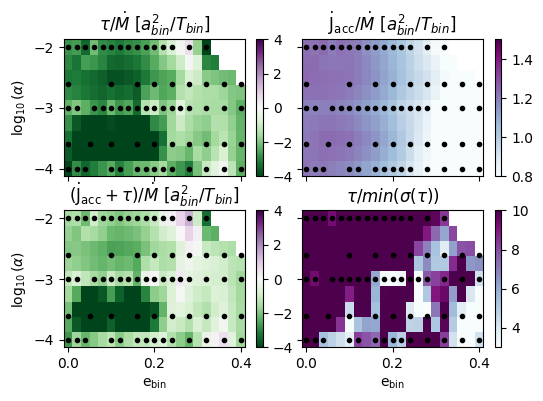}}
		\caption{The torque (top left) and the angular momentum change(bottom left) caused by the disc on the binary. The angular momentum flow normalized with massflow (top right) and gravitational power (bottom right) for different $\alpha$. Black dots mark different simulations.}
		\label{fig:sur_a_j}
	\end{figure}

Changing the viscosity through the viscous $\alpha$ coefficient does not lead to a clear trend in contrast to the aspect ratio. Overall, the orbits shrink uniformly throughout the range of $\alpha$ in Fig.~\ref{fig:sur_a}. For the binary eccentricity, there is no clear trend for either excitation or circularisation within the $\alpha$.
While some values near $\alpha=10^{-2.5}, \eb<0.2$ are increased, this is not the case for even lower or higher viscosity. The torque data at $\alpha=10^{-2.5}, \eb<0.2$ varies relative to the mean value by $\sim 0.2$ (Fig.~\ref{fig:sur_a_j}). Hence, these averaged values have a higher uncertainty.
We see weak trends, along the range of binary eccentricity. For higher eccentricities $\eb \geq 0.15 $ (right half of top Fig.~\ref{fig:sur_a}), the magnitude of the specific energy change due to gravitational forces decreases, leading to a decreased rate of shrinkage and favouring eccentricity excitation of the disc.

The increase of excitation at higher binary eccentricity can be explained by the combination of the smoothly decreasing advective momentum for higher binary eccentricities (top right panel Fig.~\ref{fig:sur_a_j}) and more uniform torque across the whole parameter space (top left panel Fig.~\ref{fig:sur_a_j}) compared to the power $\dot{E}/\dot{M}$.
The fact that the $\alpha$-coefficient does not show a clear trend for the binary evolution, but the aspect ratio does, is highly interesting as both contribute to the viscosity. The main difference between these is that $\alpha$ only contributes to the viscosity. At the same time, the scale height is also related to the sound speed, $c_\mathrm{s}$, which affects the propagation of the spiral wakes more directly.

\cite{2023siwekII} report a change in excitation behaviour beyond $\eb>0.4$, though some of our data would support this idea, the uncertainties in our simulations are too high to be conclusive.  
In the left bottom panel in Fig.~\ref{fig:sur_a_j} \& \ref{fig:sur_h_j} we show the signal-to-noise of the torque derivation as explained above. Highly eccentric binary systems show variation up to 0.3 the mean torque and should be taken with extreme caution as the variability in the torque makes an accurate measurement of the orbital evolution challenging.  

\section{Discussion}\label{sec:discuss}


The choice to exclude the binary from the domain reduces the computational time significantly, however, it also removes the information in spin angular momentum around the binary components \citep{2021Dittmann} and might lead to artificial accretion of material that would be pushed back on an outgoing trajectory. Especially for high viscosity scenarios typically considered for SMBH, the inner region between the binary components can be included in the domain, as simulations are much shorter \citep{2019Moesta,2020Munoz,2022Dittmann}.
\cite{2021Tiede} has shown that $>95\%$ of material from inside $1~\ab$ is accreted into the mini-discs or by the stars on the binary orbital time scale when the binary is included in the domain as sinks. Recent studies \citep[e.g.][]{2020Hirsh,2020Ragusa} could include the binary for protoplanetary discs using smooth particle hydrodynamics simulations. \cite{2024KITP} recently compared simulations of circumbinary discs using 9 different simulation codes, including PLUTO, finding that simulations including and excluding the binary can reproduce the circumbinary disc well. However,  the treatment of mass accretion can affect the strength of torque from gas on orbits around one of the binary components, the spin torque.
The spin torque term can change the absolute torque on the binaries by adding a torque contribution of the circumstellar discs to both binary components, instead of just assuming the instant accretion of material inside $1~\ab$. Excising the binary leads to a higher overall torque.
However, the overall torque in excised simulation domains does not vary significantly from simulations that include the binary in the domain with accretion timescales of $1-10~\Tb$. At these time scales accretion is fast enough to allow a steady flow from the cavity through the circumstellar disc onto the stars.

Binary orbital evolution has recently become of particular interest to the SMBH community but is less studied in the context of PPDs.
\cite{2017Miranda} found conditions that lead to orbit expansion for very thick and viscous discs compared to typical PPDs ($h=\alpha=0.1$), this finding was later confirmed using grids that cover the binaries by \cite{2019Munoz, 2019Moody}.
\cite{2022Franchini} reported that an even thicker discs ($h=0.2$) can lead to shrinking binaries again. This is a non-monotonic trend as 
\cite{2020Tiede, 2022Dittmann, 2022Franchini} and \cite{2022Penzlin} showed that even at high $\alpha$, the orbits of circular binaries in thin discs ($h<0.05$) shrink, which agrees with our results.
Our study shows that the findings of thin disc ($h\leq0.04$) leading to shrinking binary orbits in \cite{2020Tiede} extend to lower viscosities using $\alpha=10^{-3}$.  
\cite{2019Munoz, 2021Zrake} \& \cite{2023siwekII} included the binary eccentricity evolution for SMBH-like systems; since the models in both studies have very thick discs ($h=0.1$), the results are not directly comparable to the systems studied here. However, in this highly viscous environment, \cite{2021Zrake} and \cite{2023siwekI} show trends of orbit contraction and eccentricity excitation that appear to be inline with the result of our study. \cite{2021Zrake} and \cite{2023siwekI} extend the binary eccentricities to values beyond $0.4$, finding an additional turning point at $\eb \sim 0.4$ from eccentricity excitation to circularization. Our simulation results for the most eccentric binaries point to the possibility of a similar transition. But with $\eb \leq 0.4$ we can not confirm the full transition to the drop off at more eccentric binary eccentricities. At those eccentricities, the precession period is $> 5000~\Tb$ and the variation in the torque contributions are very large. A more detailed study of binaries with higher eccentricities would be necessary to draw firm conclusions.

\cite{2020Ragusa} studies the orbital evolution in a PPD setting, and their results show strong variations in eccentricity due to the torques and eccentricity growth. However, the simulations differ in the number of orbits simulated and include a flaring disc height. 
We consider only a single mass ratio of q=0.26 in this study, in our previous study \citep{2022Penzlin} we found that momentum and accretion contributions do not strongly depend on the mass ratio. This suggests that the retrieved balances present may be comparable values for other mass ratios. As more mass accretes onto the secondary \citep[i.e., $\lambda > 1$][]{2014Farris, 2019Munoz, 2021Dittmann, 2023siwekI}, the limits in Fig.~\ref{fig:xi} lead to a picture of unequal binaries expanding and increasing their orbital eccentricity while growing toward more even masses. As the mass ratio increases, the expansion eventually stalls and reverses first, later transitioning to eccentricity damping rather than growth. This would apply for young or large separation of binary stars, which have thick discs due to the irradiation flaring of the discs at large distances. This might be a reason for the spread in binary eccentricities in the sample of \cite{2020Whelan}.

However, the mass accretion ratio, $\lambda$, is a subject of active research. \cite{2024Dittmann} showed that the mass accretion rate, especially near mass ratios of $q=0.1$, can be sensitive to the viscosity of the system. Further, the sink size and rate at which gas is allowed to accrete into the sink 
affects
the accretion behaviour \citep{2021Dittmann, 2022Westernacher}. In the case of the binary stars, thermal conditions around each component and the conditions in the local minidiscs around the each component also alters the accretion onto the stars \citep{2021Jordan} and the possible overflow between binary components. 
Despite these complexities, the shrinking behaviour of the binaries we observe is independent of the value of $\lambda$ because the evolution is dominated by the change in energy of the binary induced by gravitational interaction with the disc, and accretion plays a subdominant role.

All simulations in this study are 2D. \cite{2024KITP} has found a small torque reduction in 3D simulations relative to 2D simulations. All simulations assume mass and momentum conservation. In a real star formation scenario, some momentum and mass are also lost through jets and outflows, which can not be considered in 2D. 
This especially affects the angular momentum transferred through accretion to the binary components as off-plane inflow does not contribute its full momentum into the orbital plane.
In cases where the mass accretion rate is reduced, the gravitational terms gain importance, aiding the shrinking and excitation of the orbit.
This adds an additional uncertainty to the final orbital changes.

\section{Conclusions}

Using more than 100 hydrodynamical simulations, we investigate the effects a circumbinary disc can have on the orbits of the central stars in a low viscous regime comparable to protoplanetary discs. By analysing the balance of mass accretion, angular momentum transport and energy change, we constrain the possible trend of binary orbital evolution.

The simulations show that the disc precessing around the binary orbit produces a strong effect on the eccentricity and semi-major axis change of the binary, in addition to the changes in the shape of the disc itself shown in \citet{2024Penzlin}(PBN1).
The mass accretion, advected angular momentum, energy and, most notably, gravitational torque show oscillating behaviour not just on the orbital time scale but also on the precession time scale of thousands of binary orbits.
This strongly impacts the evolution of the eccentricity of the binary stars, leading to a slow oscillation of eccentricity with an amplitude of the order of $10^{-3}$ in the absolute eccentricity. This oscillation potentially makes estimates of the orbital evolution derived from shorter time-scale simulations unreliable.

Within the range of binary eccentricities explored $\eb= 0.05 - 0.4$, nearly all simulations suggest increasing binary eccentricities is possible, depending on what fraction of the mass is accreted by the primary and secondary. 

Thin discs, with an aspect ratio $h<0.05$, always cause the binary orbits to shrink. However, at higher aspect ratios, both expansion and contraction are possible, depending on the degree to which mass accretion occurs preferentially onto the primary or secondary \citep[e.g.][]{2021Tiede,2022Penzlin}. Further, we find that in binaries with eccentricity below $0.2$, the gravitational forcing is stronger, and the binaries tend to shrink. For larger binary eccentricities ($\eb={0.2,0.4}$), the gravitational terms are smaller, and expansion is more likely to be possible if the mass accretion is too strongly weighted towards the secondary \citep[see also][]{2021Zrake,2023siwekI}. Finally, we do not see clear trends with viscosity within the range of values considered here ($10^{-4} \leq \alpha \leq 10^{-2}$).

The difference in mass accretion between two binary components controls the critical values of mass, angular momentum, and energy exchange that decide the trends of the orbit's evolution \citep{2017Miranda}.
If the mass accretion is higher on the secondary component, expansion and eccentricity excitation are favoured.

In low-viscosity environments, the orbital period of the binary changes by 1\%
on time scales on the order of $10^5-10^7~\Tb$ depending on the viscosity and assuming a 1\% $M_\mathrm{bin}$ disc. The time scale for the average growth of the binary eccentricity is similar. However, due to the strong effect of the precession, eccentricity variations of $\Delta \eb = 0.001$ occur within $\sim 10^3~\Tb$.

\section*{Acknowledgements}
In memory of Willy Kley, who started and pushed this project forward right until the day he passed away in 2021. 
This and many of the last decades' works were only possible through his kind advice and guidance, and he is missed deeply.
We thank Daniel Thun for his initial contribution and help in using the GPU-Pluto version, Kees Dullemond and James Owen for helpful discussion and the referee for constructive and helpful comments and suggestions.
AP and RAB acknowledge support from the Royal Society in the form of a University Research Fellowship and Enhanced Expenses Award.
Parts of this study were funded in part by grant KL 650/26 of the German Research Foundation (DFG).
This work was performed with the support of the High Performance and Cloud Computing Group at the Zentrum f\"ur Datenverarbeitung of the University of T\"ubingen, the state of Baden-W\"urttemberg through bwHPC and the German
Research Foundation (DFG) through grant no INST 37/935-1 FUGG, and Cambridge Service for Data Driven Discovery (CSD3), part of which is operated by the University of Cambridge Research Computing on behalf of the STFC DiRAC HPC Facility (www.dirac.ac.uk). The DiRAC component of CSD3 was funded by BEIS capital funding via STFC capital grants ST/P002307/1 and ST/R002452/1 and STFC operations grant ST/R00689X/1. DiRAC is part of the National e-Infrastructure.
RPN acknowledges support for the Leverhulme Trust grant RPG-2018-418 and STFC grants ST/X000931/1 and ST/T000341/1.

\section*{Data Availability}
The data underlying this article will be shared on reasonable request to the corresponding author.

\bibliography{parameter_study}
\bibliographystyle{mnras}

\begin{appendix}\label{sec:app}
\section{Steady state discs} \label{sec:app}

\begin{figure}
		\centering
		\resizebox{\hsize}{!}{\includegraphics{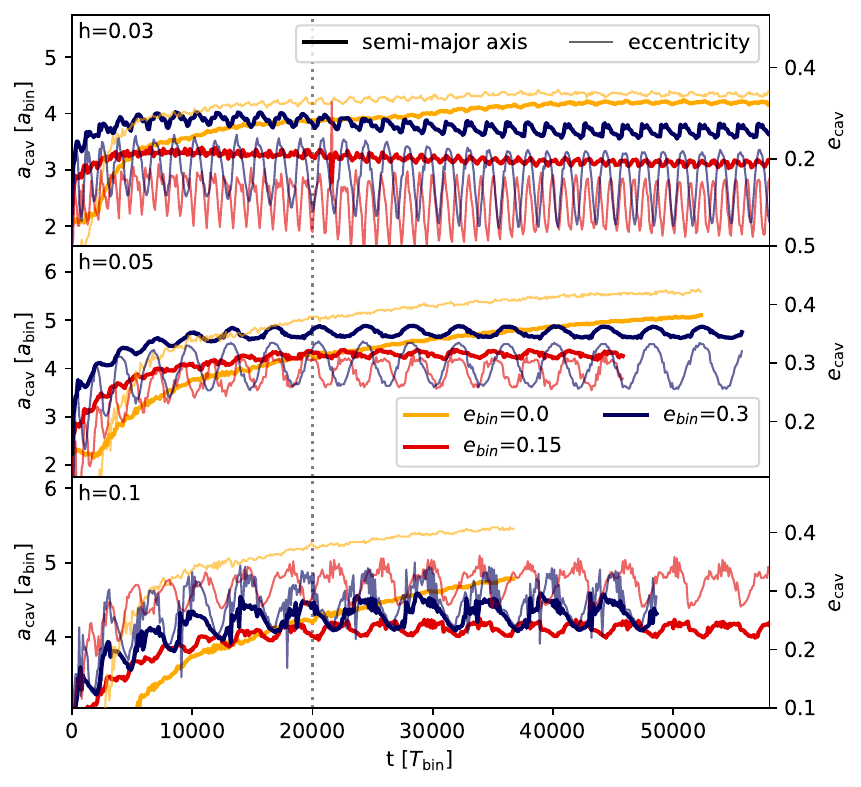}}
		\caption{Evolution of the eccentricity (thin lines) and semi-major axis (thick lines) of the steady models for different aspect ratios.}
		\label{fig:app_h_cav}
	\end{figure}

\begin{figure}
		\centering
		\resizebox{\hsize}{!}{\includegraphics{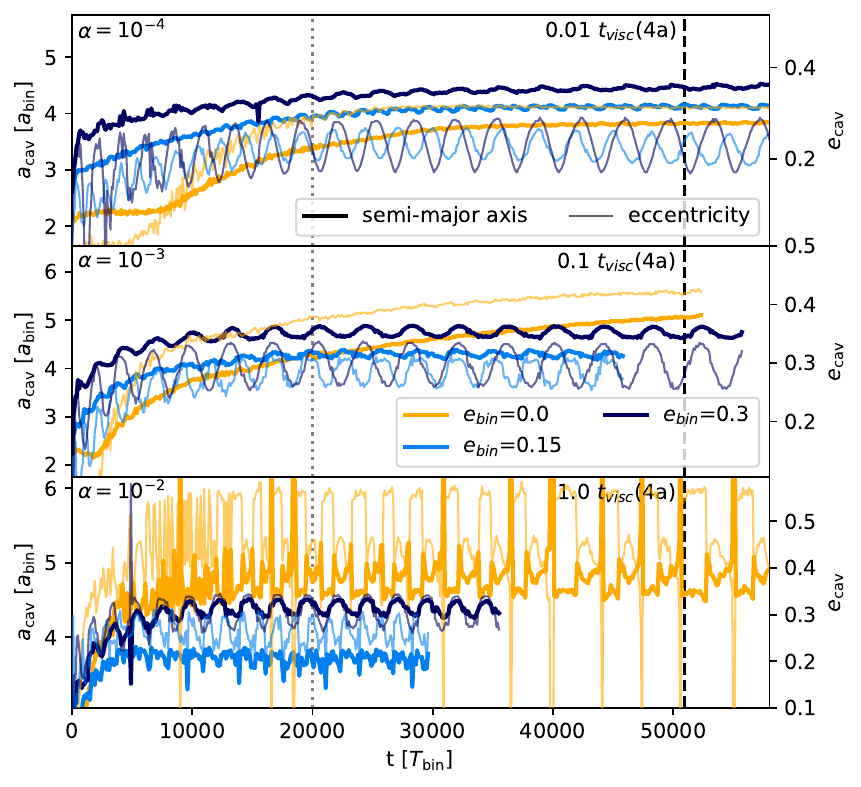}}
		\caption{Evolution of the eccentricity (thin lines) and semi-major axis (thick lines) of the steady models for different $\alpha$-values.}
		\label{fig:app_alp_cav}
	\end{figure}

In the initial setup of the study, we did not choose a surface density profile that produces a steady disc but rather a slope that is steeper than a steady condition, which acts more like a wide-spreading ring. This is evident in the mass flux in Fig.~\ref{fig:precession}. Nevertheless, the evolution of angular momentum and power is not evolving significantly in the simulations, and the only oscillation is due to precession (see also Fig.~\ref{fig:evo-time}). The discs react to the binary on timescales that differ from the viscous time, indicating that the gravitational interaction can settle quicker into its final state than on viscous times. 
To check that the non-steady initial conditions do not affect our results, we ran 15 additional steady disc models of an unflared disc with $\Sigma = \Sigma_0 R^{-1}$ 
for binary eccentricities of $\eb=[0,0.15,0.3]$, aspect ratios of $h=[0.03,0.05,0.1]$ and $\alpha=[10^{-4},10^{-3},10^{-2}$ (while the the other parameter is $h=0.05$ or $\alpha=10^{-3}$).
These models reproduce inner cavities of similar size and eccentricity as the other models, as shown in Fig.~\ref{fig:app_h_cav} \& \ref{fig:app_alp_cav} (see \cite{2024Penzlin}).
The cavity edge semi-major axes are smaller by $\leq 0.1~\ab$, as the disc provides more flow towards the binary. In addition, the models with the lowest binary eccentricity continue to rise slowly and have not reach a final state. This contrasts with the steeper density profiles that levelled off more quickly.

\begin{figure}
		\centering
		\resizebox{\hsize}{!}{\includegraphics{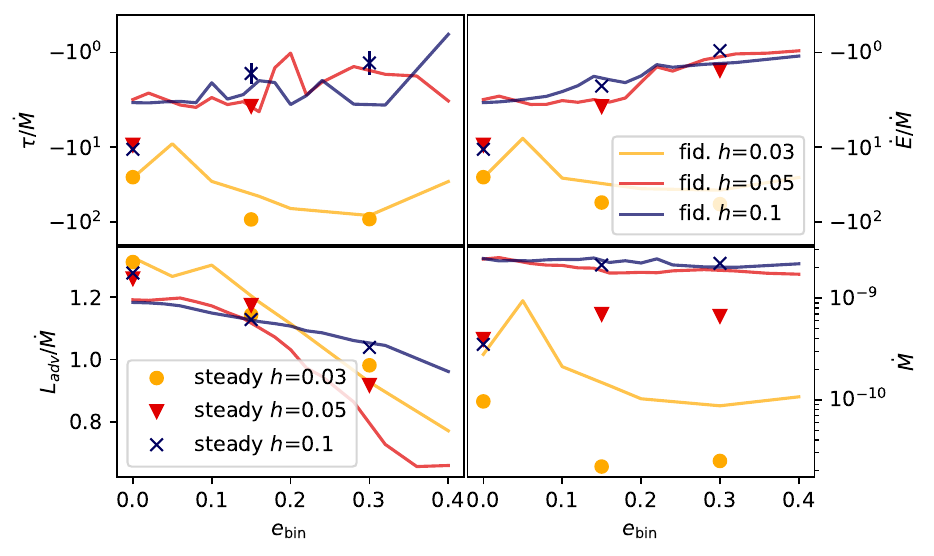}}
		\caption{Averaged gravitational torque $\tau$ (top left), gravitational power $\dot{E}$ (top left) and advected angular momentum $L_\mathrm{acc}$ (lower left), normalized by the mass accretion $\dot{M}$ (lower right). The lines represent the data off the non-steady models, the marker show the data for the steady disc solution.}
		\label{fig:app_h_tau}
	\end{figure}

\begin{figure}
		\centering
		\resizebox{\hsize}{!}{\includegraphics{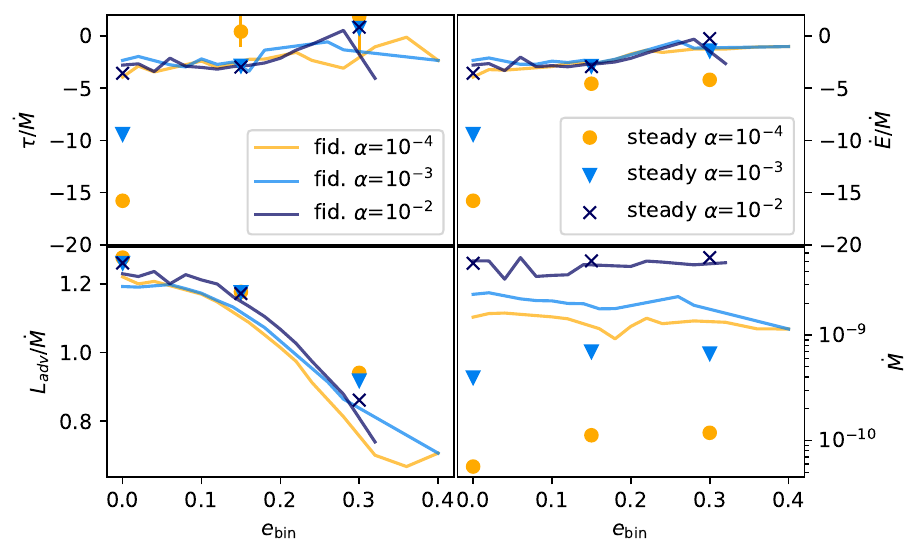}}
		\caption{Averaged gravitational torque $\tau$ (top left), gravitational power $\dot{E}$ (top left) and advected angular momentum $L_\mathrm{acc}$ (lower left), normalized by the mass accretion $\dot{M}$ (lower right). The lines represent the data off the non-steady models, the marker show the data for the steady disc solution.}
		\label{fig:app_alp_tau}
	\end{figure}

Comparing the mass normalized momentum and power of the steady simulations to the runs above, the values are similar and follow the same trend, except for the circular binaries that have not yet reached a steady cavity profile as is shown in Fig.~\ref{fig:app_h_tau} \& \ref{fig:app_alp_tau}.
For thin discs $h=0.03$, the normalized power is well above the critical value and the normalized torque is more than two times higher than other angular momentum contributions independent of the disc structure.
Hence, the shrinking and excitation of the binary orbits in such a thin disc are robust results.
The steady and non-steady models for thick discs or varying viscosity, produce similar through positive values for the normalized power. 
This indicates that orbits are likely to shrink independent of the disc model.
The steady model is in better agreement with theoretically predicted accretion rate, which are proportional the viscosity.

However, the circular binary discs deviate from this trend. This can be partly caused by the fact that they are not in an equilibrium state (yet). Still, it also shows that the excitation of the disc around an exactly circular binary is stronger, possibly due to constant interaction with the unchanging velocity profile of the gravitational wakes caused by the binary. Future studies will analyse the role of the gravitational wakes as the driver of the excitation.

\end{appendix}

\bsp	
\label{lastpage}
\end{document}